\documentclass[%
aip,
 amsmath,amssymb,
 reprint,%
]{revtex4-1}

\usepackage{graphicx}
\usepackage{dcolumn}
\usepackage{bm}

\usepackage{verbatim}
\usepackage{enumitem}
\usepackage{multirow}

\usepackage{color,soul} 

\begin{document}

	
	\title[Stress control of tensile-strained In$_{1-x}$Ga$_{x}$P nanomechanical string resonators]{Stress control of tensile-strained In$_{1-x}$Ga$_{x}$P nanomechanical string resonators}

	\author{Maximilian B\"{u}ckle}
	\affiliation{ 
		Department of Physics, University of Konstanz, D-78457 Konstanz, Germany
	}
	
	\author{Valentin C. Hauber}
	\affiliation{ 
		Department of Physics, University of Konstanz, D-78457 Konstanz, Germany
	} 
	
	\author{Garrett D. Cole}
	\affiliation{Vienna Center for Quantum Science and Technology (VCQ), Faculty of Physics, University of Vienna, A-1090 Vienna, Austria}
	
	\author{Claus G\"{a}rtner}
	\affiliation{Vienna Center for Quantum Science and Technology (VCQ), Faculty of Physics, University of Vienna, A-1090 Vienna, Austria}
	
	\author{Ute Zeimer}
	\affiliation{Ferdinand-Braun-Institut, Leibniz-Institut f\"{u}r H\"{o}chstfrequenztechnik, D-12489 Berlin, Germany}
	
	\author{J\"{o}rg Grenzer}
	\affiliation{Institute of Ion Beam Physics and Materials Research, Helmholtz-Zentrum Dresden-Rossendorf, D-01328 Dresden, Germany}
	
	\author{Eva M. Weig}
	\email{eva.weig@uni-konstanz.de}
	\affiliation{ 
		Department of Physics, University of Konstanz, D-78457 Konstanz, Germany
	}

	\date{\today}
	
	\begin{abstract}
		
		We investigate the mechanical properties of freely suspended nanostrings fabricated from tensile-stressed, crystalline In$_{1-x}$Ga$_{x}$P. 
		The intrinsic strain arises during epitaxial growth as a consequence of the lattice mismatch between the thin film and the substrate, and is confirmed by x-ray diffraction measurements. 
		The flexural eigenfrequencies of the nanomechanical string resonators reveal an orientation dependent stress with a maximum value of 650 MPa. 
		The angular dependence is explained by a combination of anisotropic Young's modulus and a change of elastic properties caused by defects. 
		As a function of the crystal orientation a stress variation of up to 50\,\% is observed. 
		This enables fine tuning of the tensile stress for any given Ga content $x$, which implies interesting prospects for the study of high Q nanomechanical systems.

	\end{abstract}

	\keywords{InGaP, GaInP, nanomechanical resonator, tensile strained crystal}
	
	\maketitle

	Introducing strain in material systems enables the control of various physical properties.
	Examples include the improved performance of semiconductor lasers,\cite{Ghiti1989ElecLett-ImprovedDynAndLinewidthEnhancementStrainedLasers,Lau1989APL-EnhancementModulationBandwidthInGaAsStrainedSQWLaser} enhanced carrier mobility in transistors,\cite{People1986IJQE-PhysicsAppsSiGeStrainedHeterostructures,Welser1994IEDL-EMobilityEnhanceStrainSiGeMOSFET} direct formation of quantum dots\cite{Leonard1993APL-DirectFormationQDsFromInGaAsOnGaAs,Landin1998Science-OpticalStudiesInAsQDsGaAs} and increased mechanical quality factors (Q) in micro- and nanomechanical systems (M-/NEMS). \cite{Verbridge2006JAP-HighQRoomTempHighStress,Faust2012NatCo-MicrowaveCavityTransduction} 
	In particular, tensile-strained amorphous silicon nitride has evolved to a standard material in nanomechanics in recent years.
	The dissipation dilution\cite{Yu2012PRL-ControlMaterialDampingHighQMembraneMicrores,GonzalezSaulson1994BrownianMotionAnelasticWire} arising from the inherent tensile prestress of the silicon nitride film gives rise to room temperature Q factors of several 100\,000 at 10\,MHz resonance frequencies,\cite{Verbridge2006JAP-HighQRoomTempHighStress,Norte2016PRL-MechResForQuantumOptomechExperimentsAtRoomtemp,Faust2012NatCo-MicrowaveCavityTransduction,Ghadimi2017NanoL-RadiationInternalLossEngineeringHighStressSiN,Villanueva2014PRL-EvidenceSurfaceLossSiN}
	while additional stress engineering has been shown to increase Q by a few orders of magnitude.\cite{Tsaturyan2017NatNano-UltracoherentNanomechSoftclamping,Norte2016PRL-MechResForQuantumOptomechExperimentsAtRoomtemp,Ghadimi2018StrainEngineering}
	However, defects\cite{Pohl2002RevModernPhys-ThermalConductivityAmorphousSolids} set a bound on the attainable dissipation and hence Q in amorphous materials,\cite{Rieviere2011PRA-OptomechSidebandCoolingMicromechOsciCloseQuantumGroundState,Faust2014PRB-TwoLevelDefectsSiN,Villanueva2014PRL-EvidenceSurfaceLossSiN} provided that other dissipation channels can be evaded.\cite{Villanueva2014PRL-EvidenceSurfaceLossSiN,Imboden2014PhysReports-DissipationNanoMechSystems}
	Stress-free single crystal resonators, on the other hand, feature lower room temperature Q factors but exhibit a strong enhancement of Q when cooled down to millikelvin temperatures,\cite{Tao2014NatCo-SingleCrystalDiamondCantilever}
	as a result of the high intrinsic Q of single crystal materials.\cite{Hamoumi2018PRL-MicroscopicNanomechDissipation}  
	Combining dissipation dilution via tensile stress with high intrinsic Q of single crystal materials could open a way to reach ultimate mechanical Q at room temperature.
	
	In the recent years, a few possible candidates for tensile-strained crystalline nanomechanical resonators have emerged. 
	Those include, for example, heterostructures of the silicon based 3C-SiC\cite{Kermany2014APL-QfactorsOverMillionStressedSiC} and the III-V semiconductors GaAs\cite{Watanabe2010APExp-FeedbackCoolingStrainedGaAs}, GaNAs\cite{Onomitsu2013APExp-UltrahighQStrainGaNAs}, and In$_{1-x}$Ga$_{x}$P.\cite{Cole2014APL-InGaP-Membrane} 
	Advantages of ternary In$_{1-x}$Ga$_{x}$P (InGaP) are the direct bandgap (for $x<63\,\%$) and the broad strain tunability.
	When grown on GaAs wafers, this alloy system may be compressively strained, strain-free or tensile strained, with possible tensile stress values exceeding 1\,GPa, by varying the group-III composition $x$. 
	The prospects of InGaP in nanomechanics range from possible applications in cavity optomechanics\cite{Cole2014APL-InGaP-Membrane,Guha2017OE-HighFreqOptomechDisk35Semicond} to coupling with quantum-electronic systems, such as quantum wells\cite{Sete2012PRA-ControlNonlinOptomechResWithQuantumWell} and quantum dots.\cite{WilsonRae2004PRL-LaserCoolingNanomechResQuantumGroundStateQuantumDot}
	
	Here we explore freely suspended nanostrings fabricated from InGaP as nanomechanical systems.
	Our analysis reveals that even for fixed $x$ the tensile stress state of the resonator can be controlled by varying the resonator orientation on the chip. 
	This implies that unlike for the case of silicon nitride NEMS, resonator orientation will be an important design parameter allowing to fine-tune the tensile stress for any given Ga content $x$.
	
\begin{figure}[th]
		\centering
		\includegraphics{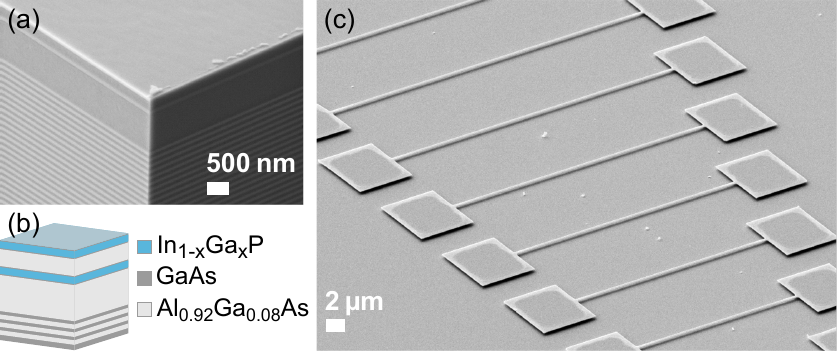}
		\caption{
			Epitaxial heterostructure. Scanning electron micrograph (a) and schematic (b) of the employed heterostructure. 
			Only the top InGaP and AlGaAs layers are used as resonator and sacrificial layer, respectively. 
			(c) String resonators with a thickness of $86\,\text{nm}$ and lengths ranging from $9\,\mu\text{m}$ to $53\,\mu\text{m}$. 
			Micrographs in (a) and (c) show high-stress InGaP.
		}
		\label{Figure1-Schematics}
	\end{figure}

	We investigate crystalline string resonators from two differently stressed, MBE grown III-V heterostructures, illustrated in Fig.\,\ref{Figure1-Schematics} (a) and (b).
	Both structures consist of two $86\,\text{nm}$ thick InGaP layers, each capped by $1\,\text{nm}$ of GaAs. 
	Both InGaP layers are situated atop a sacrificial layer of high aluminum content Al$_{y}$Ga$_{1-y}$As (AlGaAs), with $y=92\,\%$. 
	Note that only the top InGaP and AlGaAs layers were employed as resonator and sacrificial layer, respectively in this work. 
	
	By varying the Ga content of InGaP, the lattice-constant $a_\mathrm{L}^\infty(x)$ changes by up to $7\,\%$. 
	Since the substrate lattice constant of AlGaAs changes by only $0.1\,\%$, as a function of its Al content, we assume the lattice constant of AlGaAs to equal that of plain GaAs, $a_\text{AlGaAs}=a_\text{GaAs}$.
	The difference in lattice constants results in a lattice mismatch $\delta_\mathrm{L}^\infty=\left(a_\mathrm{L}^\infty(x)-a_\mathrm{GaAs}\right)/a_\mathrm{GaAs}$ between the InGaP and the GaAs lattice.
	This mismatch induces an in-plane strain $\varepsilon^\parallel(x)$ in the InGaP layer and is defined by the ratio:\cite{Pietsch2004HRXRD}
	
	\begin{align}
		\varepsilon^\parallel(x)=\frac{a_\mathrm{L}^\parallel-a_\mathrm{L}^\infty(x)}{a_\mathrm{L}^\infty(x)}~,\quad a_\mathrm{L}^\parallel=a_\text{GaAs} \label{eq:strain}
	\end{align}
	with the distorted in-plane lattice constant $a_\mathrm{L}^\parallel$ of the strained InGaP layer, which in case of a 100\,\% pseudomorphic layer equals the lattice constant of the substrate $a_\mathrm{L}^\parallel=a_\text{GaAs}$.
	An InGaP layer grows strain-free (lattice-matched) on a GaAs substrate for $x=51\,\%$ Ga content, i.e. $a_\mathrm{L}^\parallel=a_\mathrm{L}^\infty(0.51)$.\cite{Ozasa1990JAP-EffectMisfitStrainInGaP,Cole2014APL-InGaP-Membrane}
	The layer is grown tensile (compressive) strained for a higher (lower) Ga content.
	With this heterostructure it is thus possible to adjust and tailor the strain in a film up to a critical thickness determined by $x$\cite{Matthews1970JAP-AccommodationOfMisfitAcrossTheInterface,People1985APL-CalcHcVersusLatticeMismatchGeSi}.
	In this work we investigate In$_{1-x}$Ga$_{x}$P with Ga contents of $x_\text{HS} = 58.7\,\%$ (high-stress) and $x_\text{LS} = 52.8\,\%$ (low-stress).
	The resulting strain values are $\varepsilon^\parallel(x_\text{HS})=5.34\times 10^{-3}$ and $\varepsilon^\parallel(x_\text{LS})=0.95\times 10^{-3}$, for InGaP on GaAs, respectively.
	
	String resonators were defined by electron-beam-lithography followed by a SiCl$_\text{4}$ inductively coupled plasma etch, using negative electron-beam-resist \mbox{ma-N\,2403} as an etch-mask, before releasing them with a buffered HF wet etch.
	The resonators are additionally cleaned via digital wet etching.\cite{DeSalvo1996JElectrochemSoc-DigitalWetEtch}
	In the end we critical-point dried the samples, to avoid stiction and destruction of the structures.\cite{Kim1997IEEE-ReleaseMethodsCPD}
	Examples of free standing string resonators are shown in Figure\,\ref{Figure1-Schematics}\,(c).

\begin{figure}[th]
		\centering
		\includegraphics{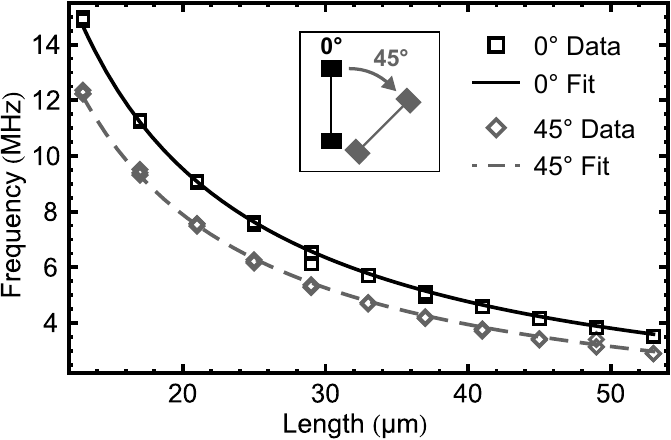}
		\caption{
			Mechanical frequencies of stressed In$_{1-x}$Ga$_{x}$P, $x_\text{HS} = 58.7\,\%$, string resonators as a function of their length, for two different orientations on the chip.
			Resonance frequencies for $0^\circ$-resonators are plotted in black rectangles and resonators rotated clockwise by $45^\circ$ in grey diamonds. 
			Fits of both datasets show the $1/L$ frequency-dependence expected for the case of strongly prestressed string resonators. 
			Calculating the weighted mean yields stress values of $\sigma(x_\text{HS},0^\circ)=642.3(3.3)$\,MPa and $\sigma(x_\text{HS},45^\circ)=440.2(2.6)$\,MPa.
			Inset: Resonator orientations with respect to chip edges. 
		}
		\label{Figure2-FreqLength}
	\end{figure}

	The samples are explored at room temperature and mounted inside a vacuum chamber (pressure $<10^{-3}$\,mbar) to avoid degradation of the AlGaAs sacrificial-layer under ambient conditions\cite{Dallesasse2013JAP-OxidationAlBearing35Review} as well as gas damping.
	We measured the fundamental resonance frequency of the out-of-plane flexural mode of resonators of different length and orientation on the substrate, using piezo-actuation and interferometric detection.
	The InGaP resonators exhibit quality factors up to 70\,000.
	Figure\,\ref{Figure2-FreqLength} presents the measured frequencies of several sets of resonators fabricated from the high-stress InGaP epitaxial structure as a function of the resonator length $L$ for two different resonator orientations on the chip.
	Resonators with an angle of $0^\circ$ are oriented parallel to the cleaved chip edges, see inset of Fig.\,\ref{Figure2-FreqLength}, which correspond to the $<$110$>$ crystal directions for III-V heterostructures on (001) GaAs substrate wafers.
	Hence, the strings point along a $<$110$>$ direction.
	For comparison, we also discuss resonators which are rotated clockwise by $45^\circ$, and hence are oriented along a $<$100$>$ direction of the crystal.
	
	Following Euler-Bernoulli beam theory,\cite{Timoshenko1990-VibrationProblemsEngineering,Cleland2002foundations} we can express the eigenfrequency of the $n$-th harmonic as
	\begin{subequations}
		\begin{align}
			f_n &= \frac{n^2 \pi}{2 L^2} \sqrt{\frac{E I}{\rho A}}\sqrt{1+ \frac{\sigma A L^2}{n^2 \pi^2 E I}} \label{eq:frequencyA}\\
			f_n &\approx  \frac{n}{2 L}\sqrt{\frac{\sigma}{\rho}}~~\text{for}~~\frac{\sigma A L^2}{n^2 \pi^2 E I} \gg 1, \label{eq:frequencyB}
		\end{align}
	\end{subequations}
	where $E$ is the Young's modulus, $I$ is the area moment of inertia, $\rho$ is the mass density, $A$ is the cross-sectional area, and $\sigma$ the stress.
	For the case of sufficiently strong tensile stress Eq.\,\ref{eq:frequencyA} reduces to Eq.\,\ref{eq:frequencyB}.
	The resonance frequencies shown in Fig.\,\ref{Figure2-FreqLength} are fitted with Eq.\,\ref{eq:frequencyB} and clearly follow the expected $1/L$ dependence.
	Being in the high tensile stress regime, a change in frequency for a given resonator length can only originate from a different tensile stress $\sigma$.
	The frequency mismatch between the $0^\circ$ and $45^\circ$ data indicates that the stress depends on the resonator's orientation.
	Solving Eq.\,\ref{eq:frequencyA} for $\sigma$ and calculating the weighted mean from all data points yields $\sigma(x_\text{HS},0^\circ)=642.3(3.3)$\,MPa and $\sigma(x_\text{HS},45^\circ)=440.2(2.6)$\,MPa, indicating that the tensile stress varies by almost $50\,\%$ with crystal direction.

	For anisotropic materials, stress $\sigma$ and strain $\varepsilon$ are related by the fourth rank compliance $S$ or stiffness $C$ tensors, $\sigma = C \varepsilon$ and $\varepsilon = S \sigma$.\cite{Hopcroft2010JMEMS-WhatYoungsModulusSilicon}
	For cubic crystals, those tensors simplify to $6\times6$ matrices with three independent components, $c_{11}$, $c_{12}$ and $c_{44}$ (see \hyperlink{appendix1}{supplementary material}). 
	For In$_{1-x}$Ga$_{x}$P, each component $c_\text{ij}(x)$ depends on the Ga content $x$, values are taken from Ref.\,\onlinecite{Ioffe1999ShurEtAl-HandbookSeriesSemiconductorParametersVOL2}.
	
\begin{figure}[th!]
		\centering
		\includegraphics{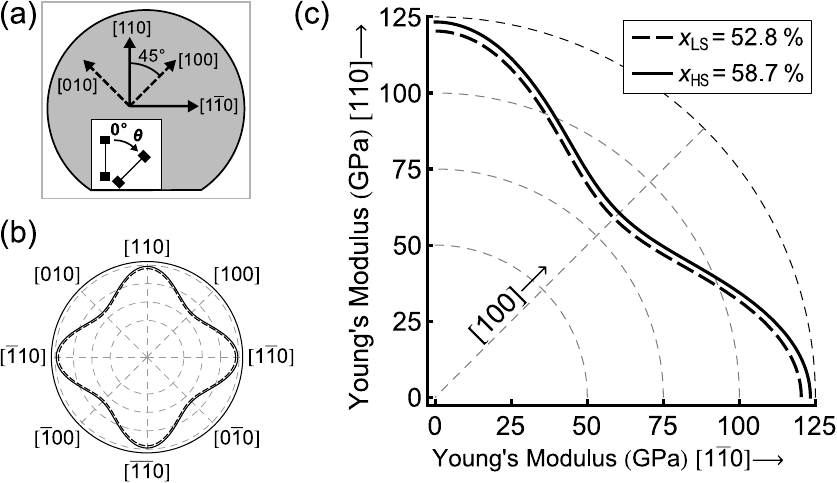}
		\caption{
			Crystal orientation in wafers and angle dependent Young's modulus in In$_{1-x}$Ga$_{x}$P. 
			(a) Schematic crystal orientations of a (001) GaAs wafer. 
			In this case, unrotated ($0^\circ$) resonators point along a $<$110$>$ crystal direction. 
			The resonator angles are changed clockwise, e.g. from [110] towards [100]. 
			Inset: Definition of the resonator angle such that $0^\circ$ resonators are parallel to the chip edge along a $<$110$>$ direction. 
			(b) Orientation dependent Young's modulus inside the (001) wafer plane, showing a $90^\circ$ rotation symmetry. 
			Solid line for  $x_\text{HS}= 58.7\,\%$ and dashed line for $x_\text{LS}=52.8\,\%$. 
			(c) Close-up of angle dependent Young's modulus, showing the first quadrant of the polar plot (b). 
		}
		\label{Figure3-YoungsModulus}
	\end{figure}

	By applying matrix rotations and transformations, one can calculate the angle dependent Young's modulus $E(x,\theta)$ of an ideal and defect free system (see \hyperlink{appendix1}{supplementary material}). 
	Figure\,\ref{Figure3-YoungsModulus} shows $E(x,\theta)$ for the two different Ga contents $x_\text{HS} = 58.7\,\%$ and $x_\text{LS} = 52.8\,\%$. 
	The Young's modulus displays a similar behavior for both Ga contents, and varies between 80\,GPa and 125\,GPa, between the $<$100$>$ and $<$110$>$ crystal directions, respectively. 
	In addition, Fig.\,\ref{Figure3-YoungsModulus} clearly reveals the $90^\circ$ rotation symmetry of $E(x,\theta)$.
	To calculate the tensile stress we multiply the Young's modulus by the strain from Eq.\,\ref{eq:strain} according to Hooke's law:
	\begin{align}
		\sigma(x,\theta) = E(x,\theta) \varepsilon^\parallel(x)~. \label{eq:StressStrainHooke}
	\end{align}
	The resulting stress values for both angles, $\sigma(x_\text{HS},0^\circ)= 655.3\,\text{MPa}$ and $\sigma(x_\text{HS},45^\circ)= 454.9\,\text{MPa}$, coincide well with the experimental results.
	
\begin{figure}[th!]
		\centering
		\includegraphics{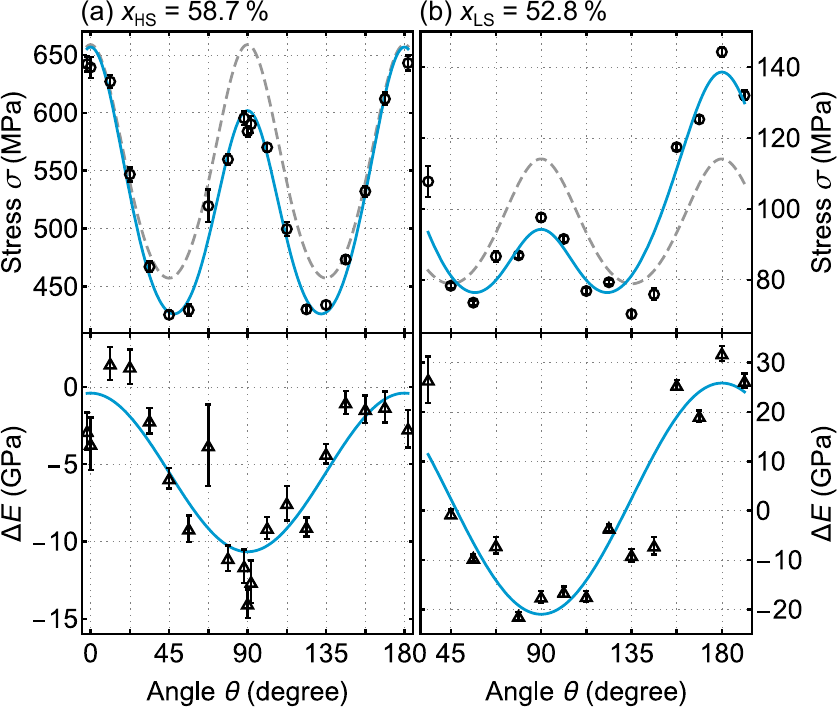}
		\caption{
			Angular stress dependence of tensile strained In$_{1-x}$Ga$_{x}$P string resonators. 
			(a) High-stress InGaP with Ga content of $x_\text{HS}=58.7\,\%$. 
			Stress varying between $430\,\text{MPa}$ and $640\,\text{MPa}$ (top). 
			Dashed gray line: Theoretically calculated stress , using Eqs. \ref{eq:strain} and \ref{eq:StressStrainHooke}. 
			Blue line: Taking a change of elastic properties due to defects into account by a $\cos(2\theta)$ angle dependent change $\Delta E$  of the Young's modulus $E(x,\theta)$ (bottom). 
			(b) Low-stress InGaP with $x_\text{LS}=52.8\,\%$. 
			Showing a similar change of the Young's modulus as in (a).
			Error bars represent the uncertainty from the weighted mean calculation.
		}
		\label{Figure4-AngularStress}
	\end{figure}

	To further investigate the angular stress dependence of InGaP, we fabricated similar sets of resonators with angles changing in $\Delta\theta=11.25^\circ$ steps. 
	For each orientation the tensile stress is extracted using Eq.\,\ref{eq:frequencyA}.
	The top plots of Fig.\,\ref{Figure4-AngularStress} show the resulting angular stress dependence for two different Ga contents. 
	In both cases, local stress maxima are observed at $0^\circ$ and $90^\circ$, i.\,e. along $<$110$>$ crystal directions. 
	Accordingly, the minima are found at $45^\circ$ and $135^\circ$, which correspond to $<$100$>$ directions.
	
	The gray dashed line in Fig.\,\ref{Figure4-AngularStress} depicts the stress values obtained by using Eqs. \ref{eq:strain} and \ref{eq:StressStrainHooke}, which does not completely coincide with the experimental data.
	While the model conforms with the data at $0^\circ$ and $180^\circ$, there are deviations around $90^\circ$  for both $x_\text{HS}=58.7\,\%$ and $x_\text{LS}=52.8\,\%$. 
	
\begin{figure}[th!]
		\centering
		\includegraphics{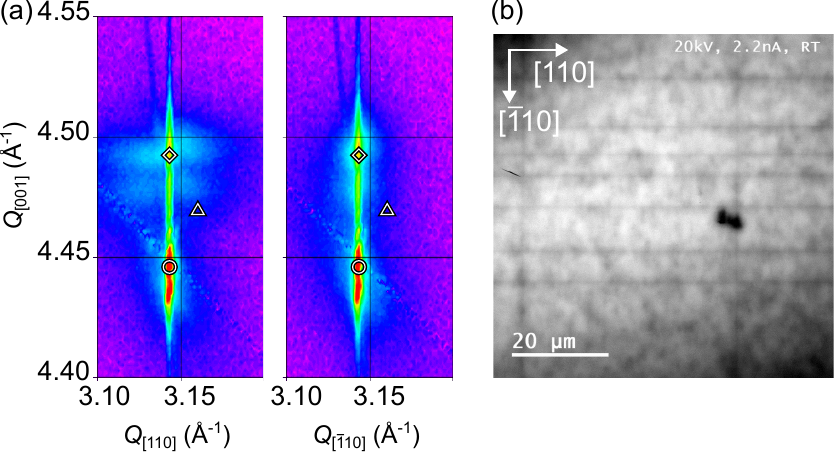}
		\caption{
			(a) Reciprocal space maps depicting the asymmetric 224 reflections of the HRXRD measurement.
			On the left the impinging x-ray beam is oriented along the [110] and on the right along the $[\overline{1}10]$ direction.
			$Q_\mathrm{[hkl]}$ are the reciprocal lattice vectors. 
			The circle indicates the substrate peak. 
			The layer peak position for a 100\,\% pseudomorphic InGaP layer is indicated by a diamond.
			In contrast, the triangle indicates the position of a fully relaxed layer.
			The coincidence of the observed layer peak with the diamond confirms that the unstructured InGaP is 100\,\% pseudomorphic.
			(b) Cathodoluminescence measurements are used to elucidate the dislocation density in the strained epitaxial structure. As shown in this image, the orthogonal $<$110$>$ crystal directions exhibit a different density of dislocation lines (horizontal and vertical dark lines), resulting in variations of the defect structure as a function of orientation.
		}
		\label{Figure5-XRDCL}
	\end{figure}

	To elucidate the deviation of stress, we have performed high resolution x-ray diffraction (HRXRD) reciprocal space map measurements\cite{Pietsch2004HRXRD} along two orthogonal $<$110$>$ crystal directions as shown in Fig.\,\ref{Figure5-XRDCL}\,(a).
	The diffraction peak arising from the InGaP layer lies directly above the substrate peak (circles in Fig.\,\ref{Figure5-XRDCL}(a)), i.e. at the same $Q_{<110>}$ positions, and coincides with the expectation for a 100\,\% pseudomorphic film within an accuracy of $10^{-3}\,\text{\AA}^{-1}$. 
	In particular the HRXRD measurements show the same out-of-plane strain for both the $[110]$ and $[\overline{1}10]$ sample orientations.
	However, the InGaP layer peaks show a different diffuse scattering which can be mainly attributed to point defects,\cite{Kaganer1997-XrayDiffPeaksDislocations} indicating different defect densities along the orthogonal $<$110$>$ crystal directions.
	
	Additional cathodoluminescence (CL) measurements in Fig.\,\ref{Figure5-XRDCL} (b) are done to obtain further insight on the dislocation density in the epitaxial material. 
	Dislocation lines have a higher density along the $[\overline{1}10]$ direction than along $[110]$. 
	These measurements confirm different defect densities along the orthogonal $<$110$>$ crystal directions.  
	
	It has been shown, that defects can influence the elastic properties of crystalline materials, and can lead to a softening as well as a hardening of the elastic constants.\cite{Dai2015NanoLett-ElasticPropGaNNanowireDefectYoungsModulus,Chen2016NanoLett-GaAsNanowiresStackingFaultsYoungsModulus}
	
	This change of elastic properties can be treated as an effective Young's modulus $\sigma(x,\theta)/\varepsilon^\parallel(x) = E(x,\theta)+\Delta E(\theta)$.
	We extract the deviation $\Delta E(\theta)$ from the experimentally obtained stress, the strain using Eq.\,\ref{eq:strain}, and the theoretically calculated Young's modulus determined in Fig.\,\ref{Figure3-YoungsModulus}.
	The extracted  values are shown in the bottom plots of Fig.\,\ref{Figure4-AngularStress} and clearly reveal an angular deviation from the theoretical Young's modulus.
	Both the softening and hardening of elastic constants can be seen for our two different InGaP compositions. 
	One can see softening for the high-stress sample, while the low-stress sample shows both softening and hardening. 
	Fitting a phenomenological $\cos(2\theta)$ function to the data, leads to the deviation functions $\Delta E_\text{HS}(\theta) = \left(-5.53 + 5.13\cos(2\theta)\right)$\,GPa	and $\Delta E_\text{LS}(\theta) = \left(2.44 + 23.40\cos(2\theta)\right)$\,GPa, respectively.
	Adding those functions to the theoretical Young's modulus to calculate the angular stress (Eq. \ref{eq:StressStrainHooke}), we obtain the solid blue lines in Fig.\,\ref{Figure4-AngularStress} which indicate the added effect.

	In conclusion, we have explored tensile-strained nanomechanical string resonators fabricated from crystalline In$_{1-x}$Ga$_{x}$P. 
	The initial InGaP thin film is pseudomorphically strained for a thickness of $86\,\text{nm}$ and a Ga content of $x_\text{HS} = 58.7\,\%$. 
	For the given composition we extracted an angle-dependent tensile stress of up to 650\,MPa.
	InGaP with a Ga content of $x_\text{LS}=52.8\,\%$ shows lower tensile stress around 100\,MPa with a similar angle-dependence as the high-stress InGaP.
	The observed angular stress dependence with respect to the crystal orientation is explained by a combination of anisotropic Young's modulus and a change of elastic properties caused by defects.
	This enables control over the stress of a nanomechanical resonator for a given heterostructure with fixed Ga content, which in turn could be optimized to enable maximum tensile stress. 
	In addition, angular stress control opens a way to investigate the influence of tensile stress on the dissipation of nanomechanical systems.
	Stress control and further characterization of strained crystalline resonators will help to gain a deeper understanding in pursuit of ultimate mechanical quality factors.\cite{Tsaturyan2017NatNano-UltracoherentNanomechSoftclamping,Ghadimi2018StrainEngineering}
	Finally, InGaP is a promising material for cavity optomechanics, as two photon absorption is completely suppressed at telecom wavelengths.\cite{Cole2014APL-InGaP-Membrane,Guha2017OE-HighFreqOptomechDisk35Semicond}
	Moreover, tensile strained InGaP could open a way to combine a high Q nanomechanical system with a quantum photonic integrated circuit on a single chip.\cite{Dietrich2016-GaAsIntegratedQuantumPhotonics,Bogdanov2017OME-MaterialPlatformsIntegratedQuantumPhotonics}
	\\
	
	\hypertarget{appendix1}{ }
	See \hyperlink{appendix2}{supplementary material} for detailed descriptions of the fabrication process, calculations of the Young's modulus, comments on the critical thickness of InGaP lattice matched to GaAs, and for more details on the HRXRD measurements.
	\\

	G.D.C. would like to thank Chris Santana and the team at IQE NC for the growth of the epitaxial material used in this study.
	Financial support by the Deutsche Forschungsgemeinschaft via the collaborative research center SFB\,767, the European Union’s Horizon 2020 Research and Innovation Programme under Grant Agreement No 732894 (FET Proactive HOT), and the German Federal Ministry of Education and Research (contract no. 13N14777) within the European QuantERA co-fund project QuaSeRT is gratefully acknowledged. 	
	
	The data and analysis code used to produce the nanomechanical plots are available at \url{http://dx.doi.org/10.5281/zenodo.1477912}.

	%


\newpage

\clearpage

\appendix

\numberwithin{equation}{section}
\numberwithin{figure}{section}
\numberwithin{table}{section}

\begin{widetext}

	\setcounter{figure}{0}
	\setcounter{equation}{0}
	
	\renewcommand{\thepage}{S\arabic{page}} 
	\renewcommand{\thesubsection}{S\arabic{subsection}} 
	\renewcommand{\thesubsubsection}{\alph{subsubsection}}
	\renewcommand{\thetable}{S\arabic{table}}  
	\renewcommand{\thefigure}{S\arabic{figure}}
	\renewcommand{\theequation}{S\arabic{equation}}
	
	
	
	\hypertarget{appendix2}{ }
	
	\clearpage
	
	\section*{Supplementary material:\\ S\lowercase{tress control of tensile-strained} I\lowercase{n}$_{1-x}$G\lowercase{a}$_{x}$P \lowercase{nanomechanical string resonators}}

	\subsection{Fabrication}
	
	The employed heterostructures were grown by molecular beam epitaxy on  $150\,\text{mm}$ diameter, $675\,\mu\text{m}$ thick (001) GaAs wafers. 
	Both In$_{1-x}$Ga$_{x}$P (InGaP) layers on each wafer  have a thickness of 86\,nm, and are capped by 1\,nm GaAs on both sides, respectively.
	The two investigated Ga contents are $x_\text{HS} = 58.7\,\%$ (high-stress) and $x_\text{LS} = 52.8\,\%$ (low-stress).\\
	From bottom to top, the structure consists of a GaAs buffer layer followed by a GaAs/Al$_y$Ga$_{1-y}$As distributed Bragg reflector.
	The bottom sacrificial Al$_y$Ga$_{1-y}$As (AlGaAs) layer has a thickness of 1065\,nm.
	The following two InGaP layers are separated by the 265\,nm thick top sacrificial Al$_y$Ga$_{1-y}$As layer.
	The aluminum content of all Al$_y$Ga$_{1-y}$As layers is $y=92\,\%$.
	We only use the top InGaP and sacrificial AlGaAs in this work.
	
	String resonators are defined by electron-beam-lithography.
	The negative electron-beam-resist \mbox{ma-N\,2403} serves as etch-mask.
	To avoid delamination, we apply an adhesion promoter TI Prime before spin-coating the resist.
	The roughly 240\,nm thick resist features good resistance against dry and wet etches.
	Etch-resistance is further increased by an additional hard bake after development, for 10\,min at 120$^\circ$C in a convection oven.
	We pattern the string resonators with an inductively coupled plasma (ICP) etch, using a SiCl$_4$:Ar (1:3) gas mixture.
	Etching is carried out at a pressure of 1.7\,mTorr with 250\,W ICP power and 60\,W RF power and at a temperature of 30$^\circ$C.
	To remove possible chlorine residues of the ICP etch, we soak the samples for 10\,min in DI water.
	With an oxygen plasma cleaner we remove the resist etch-mask.
	The following buffered HF etch releases the string resonators, by etching the sacrificial AlGaAs with an etch rate of 50--90\,nm/s, depending on the crystal direction.
	After thoroughly rinsing the samples, we use a digital wet etch\cite{DeSalvo1996JElectrochemSoc-DigitalWetEtch} to remove the GaAs cap layers and possible etch residues.
	In the end we dry the samples via critical point drying.
	To avoid degradation of the AlGaAs sacrificial-layer under ambient conditions,\cite{Dallesasse2013JAP-OxidationAlBearing35Review} the samples are quickly mounted inside the vacuum chamber of the measurement setup.
	
	Under ambient conditions the AlGaAs sacrificial layer quickly degrades and swells, 
	fracturing the suspended InGaP resonators, see Fig.\,\ref{SupplementFigure1-RTO}\,(a),\,(b).
	Alternatively the AlGaAs surface can be passivated by rapid thermal oxidation (RTO) as shown in Fig.\,\ref{SupplementFigure1-RTO}\,(c)-(e).
	This is done by a 5\,min long rapid thermal anneal at 550$^\circ$C in an oxygen atmosphere.
	The AlGaAs surface is stable for at least three weeks.
	To what extent this treatment changes the mechanical properties of InGaP string resonators, remains a topic of further investigation.

	\begin{figure}[th]
		\centering
		\includegraphics{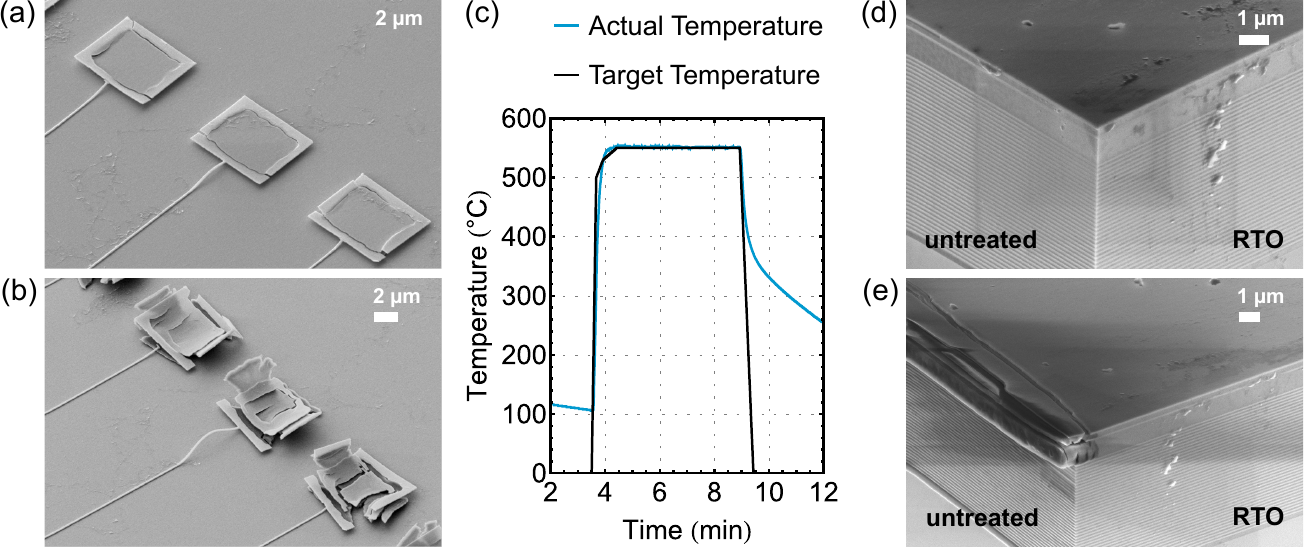}
		\caption{
			AlGaAs degradation and rapid thermal oxidation (RTO). 
			(a) InGaP nanoresonators after a few days in ambient air. 
			Cracks at the InGaP clamping points are already visible due to the degradation of the underlying AlGaAs.
			(b) The same clamping points as in the image above are completely destroyed after 8 weeks in ambient air.
			(c) Temperature profile of the 5\,min long RTO, with a 100\,sccm O$_2$ flow. 
			(d) Scanning electron micrograph of the heterostructure, right surface treated with RTO and left untreated side produced by a fresh cleave right before imaging.
			(e) Heterostructure after 21 days in ambient conditions. 
			Untreated surface shows severe degradation from swollen AlGaAs layers while the treated surface remains unaffected. 
		}
		\label{SupplementFigure1-RTO}
	\end{figure}

	\clearpage
	\subsection{Calculating Young's modulus}
	
	The Young's modulus relates stress and strain in the one-dimensional case of isotropic, uniaxial materials via Hooke's law $\sigma = E \varepsilon$.
	For anisotropic materials, stress and strain are related by the fourth rank compliance $S$ or stiffness $C$ tensors, $\sigma = C \varepsilon$ and $\varepsilon = S \sigma$.\cite{Hopcroft2010JMEMS-WhatYoungsModulusSilicon}
	Those tensors are simplified to $6\times6$ matrices with three independent components, for the case of the cubic symmetry of e.g. the zincblende crystal structure.
	For the [100] crystal direction, Young's modulus simply equals the inverse first component of the compliance matrix $S$:
	\begin{align}
		s_{11} = \frac{1}{E}~. \label{eqAppendix:s11E100}
	\end{align}
	However, since we know the elastic constants $c_\text{ij}(x)$ of In$_{1-x}$Ga$_{x}$P,\cite{Ioffe1999ShurEtAl-HandbookSeriesSemiconductorParametersVOL2} we start with the stiffness matrix to calculate the Young's modulus for any desired crystal direction:
	\begin{align}
		C(x) = \left(\begin{matrix}
			c_{11}(x) & c_{12}(x) & c_{12}(x) & 0 & 0 & 0 \\ 
			c_{12}(x) & c_{11}(x) & c_{12}(x) & 0 & 0 & 0 \\ 
			c_{12}(x) & c_{12}(x) & c_{11}(x) & 0 & 0 & 0 \\ 
			0 & 0 & 0 & c_{44}(x) & 0 & 0 \\ 
			0 & 0 & 0 & 0 & c_{44}(x) & 0 \\ 
			0 & 0 & 0 & 0 & 0 & c_{44}(x)
		\end{matrix}\right) ~.
		\label{eqAppendix:CxMatrix}
	\end{align}
	Using the rotation-matrices Eqs.\,\ref{eqAppendix:MatrixRotations}, we can perform clockwise rotations through an angle $\theta$ about a desired major crystal axis, initially $X$=[100], $Y$=[010] and $Z$=[001]:
	\begin{align}
		C(x,\theta) = \big(\text{Rot}_\text{i}(\theta)\cdots\text{Rot}_\text{j}(\theta)~ C(x)\big)\text{Rot}^\top_\text{j}\cdots\text{Rot}^\top_\text{i}~,~\text{i, j}=X, Y, Z~.
	\end{align}
	For example the application of $\text{Rot}_\text{Z}(45^\circ)$ to a [100] direction, produces a vector pointing along the [110] direction. 
	By inverting the rotated stiffness matrix, we calculate the compliance matrix:
	\begin{align}
		S(x,\theta) = \big( C(x,\theta) \big) ^{-1}~.
	\end{align}
	As in Eq. \ref{eqAppendix:s11E100}, the Young's modulus is the inverted first component $s_{11}$ of the compliance matrix. But now it is along the rotated [100] direction, thus pointing in any desired direction:
	\begin{align}
		E(x,\theta) = s_{11}^{-1}(x,\theta) 
	\end{align}
	For the Young's modulus in the (001) plane, it simplifies to the following equation:
	\begin{align}
		E(x,\theta)\mathord{=}\frac{8 \big(c_{11}(x) \mathord{-} c_{12}(x)\big) \big(c_{11}(x) \mathord{+} 2 c_{12}(x)\big) c_{44}(x)}{c_{11}^2(x)\mathord{-}2c_{12}(x) \big(c_{12}(x) \mathord{-} 2 c_{44}(x)\big) \mathord{+} c_{11}(x) \big(c_{12}(x) \mathord{+} 6 c_{44}(x)\big) \mathord{+} \big(c_{11}(x) \mathord{+} 2 c_{12}(x)\big) \big(c_{11}(x) \mathord{-} c_{12}(x) \mathord{-} 2 c_{44}(x)\big) \cos(4 \theta)}
		\label{eqAppendix:Extheta}
	\end{align}
	which is used to calculate the contour lines in Fig.\,3 of the main paper.
	A full three dimensional plot of the Young's modulus is shown in Fig.\,\ref{SupplementFigure2-YoungsMod3D}. 
	The red line represents Eq.\,\ref{eqAppendix:Extheta}, the area cut through the (001) plane.	
	
	\begin{subequations}
		\label{eqAppendix:MatrixRotations}
		\begin{align}
			\text{Rot}_\text{X}(\theta) = \left(\begin{matrix}
				1 & 0 & 0 & 0 & 0 & 0 \\ 
				0 & \cos^2(\theta) & \sin^2(\theta) & \sin(2\theta) & 0 & 0 \\ 
				0 & \sin^2(\theta) & \cos^2(\theta) & -\sin(2\theta) & 0 & 0 \\ 
				0 &  -\frac{1}{2}\sin(2\theta) & \frac{1}{2}\sin(2\theta) & \cos(2\theta) & 0 & 0 \\ 
				0 & 0 & 0 & 0 & \cos(\theta) & -\sin(\theta) \\ 
				0 & 0 & 0 & 0 & \sin(\theta) & \cos(\theta)
			\end{matrix}  \right)
			\\
			\text{Rot}_\text{Y}(\theta) = \left(\begin{matrix}
				\cos^2(\theta) & 0 & \sin^2(\theta) & 0 & -\sin(2\theta) & 0 \\ 
				0 & 1 & 0 & 0 & 0 & 0 \\ 
				\sin^2(\theta) & 0 & \cos^2(\theta) & 0 & \sin(2\theta) & 0 \\ 
				0 &  0 & 0 & \cos(\theta) & 0 & \sin(\theta) \\ 
				\frac{1}{2}\sin(2\theta) & 0 & -\frac{1}{2}\sin(2\theta) & 0 & \cos(2\theta) & 0 \\ 
				0 & 0 & 0 & -\sin(\theta) & 0 & \cos(\theta)
			\end{matrix}  \right)
			\\
			\text{Rot}_\text{Z}(\theta) = \left(\begin{matrix}
				\cos^2(\theta) & \sin^2(\theta) & 0 & 0 & 0 & \sin(2\theta) \\ 
				\sin^2(\theta) & \cos^2(\theta) & 0 & 0 & 0 & -\sin(2\theta) \\ 
				0 & 0 & 1 & 0 & 0 & 0 \\ 
				0 & 0 & 0 & \cos(\theta) & -\sin(\theta) & 0 \\ 
				0 & 0 & 0 & \sin(\theta) & \cos(\theta) & 0 \\ 
				-\frac{1}{2}\sin(2\theta) & \frac{1}{2}\sin(2\theta) & 0 & 0 & 0 & \cos(2\theta)
			\end{matrix}  \right)
		\end{align}
	\end{subequations}
	
	\begin{figure}[th!]
		\centering
		\includegraphics{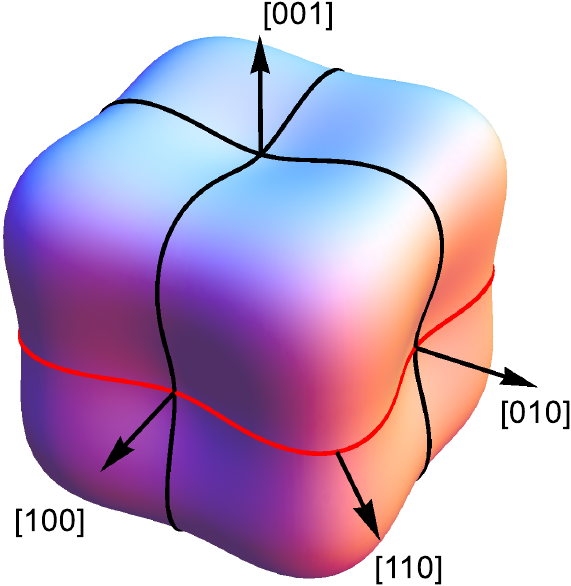}
		\caption{3D plot of Young's modulus of InGaP. Arrows indicate major crystal directions. Red line represents the cut through the (001) plane as depicted in Fig.\,3 of the main paper.}
		\label{SupplementFigure2-YoungsMod3D}
	\end{figure}

	\clearpage
	\subsection{Critical thickness}
	
	This section provides an overview of the existing models to calculate the critical thickness of strained epilayers, and is a summary of Refs. \onlinecite{Matthews1970JAP-AccommodationOfMisfitAcrossTheInterface,People1985APL-CalcHcVersusLatticeMismatchGeSi}. 
	
	There are two different approaches to treat the generation of dislocations, and thus yield a different thickness when a strained epilayer starts to relax.
	The first approach is the force-balancing model by Matthews.\cite{Matthews1970JAP-AccommodationOfMisfitAcrossTheInterface} 
	This model considers the forces on dislocation lines.
	Those are misfit strain (as driving force), which is opposed by the tension in the misfit dislocation line.
	An epilayer starts to relax when the force exerted by the misfit strain becomes larger than the tension in a dislocation line.
	The other approach to calculate the critical thickness is the energy-balancing model by People and Bean.\cite{People1985APL-CalcHcVersusLatticeMismatchGeSi}
	It compares the homogeneous strain energy density with the energy density associated with the generation of dislocations.
	If the surface strain energy density exceeds the self-energy of an isolated dislocation, dislocations are introduced which lead to a relaxation.
	
	The formulas to calculate the critical thickness $h_\text{c}$ are as follows:
	\begin{itemize}
		\item Matthews:
		\begin{align}
			h_\text{c} = \frac{b}{4 \pi \varepsilon^\parallel}\frac{(1-\nu \cos^2(\Theta))}{ (1+\nu) \cos(\alpha)} \ln\left(\frac{h_\text{c}}{b}\right)
		\end{align}
		
		\item People \& Bean:
		\begin{align}
			h_\text{c} = \left(\frac{1-\nu}{1+\nu}\right) \frac{b^2}{16\pi\sqrt{2} \,a_\mathrm{L}^\infty\, \varepsilon^{\parallel2}}  \ln\left(\frac{h_\text{c}}{b}\right)
		\end{align}
		
	\end{itemize}
	
	\begin{itemize}[leftmargin=*]
		\item[] $a_\mathrm{L}^\infty$: lattice constant of In$_{1-x}$Ga$_{x}$P
		\item[] $b=a_\mathrm{L}^\infty/\sqrt{2}$: magnitude of Burgers vector of dislocation
		\item[] $\nu= c_{12}/(c_{11}+c_{12})$: Poisson ratio of InGaP; $c_\text{ij}$ elastic constants.
		\item[] $\varepsilon^\parallel(x)=\frac{a_\mathrm{L}^\parallel-a_\mathrm{L}^\infty(x)}{a_\mathrm{L}^\infty(x)}$: in-plane strain, $a_\mathrm{L}^\parallel$ distorted lattice constant due to lattice mismatch, as in Eq.\,1.
		\item[] $\Theta$: angle between dislocation line and Burgers vector (60$^\circ$ for most III-V semiconductors)
		\item[] $\alpha$: angle between slip direction and direction in epilayer plane which is perpendicular to the line of intersection of the slip plane and the interface (60$^\circ$) 
	\end{itemize}
	By solving these formulas we obtain the curves of Fig.\,\ref{SupplementFigure3-CriticalThickness}, which demonstrate that the model of Matthews gives a more conservative estimate of the critical thickness than the one by People \& Bean. 
	The horizontal line marks the thickness of the investigated InGaP film, whereas the vertical lines indicate the two Ga contents. 
	Clearly, the model of Matthews is not able to describe the pseudomorphic high stress material which would greatly exceed the critical thickness.
	We thus conclude that the MBE growth process is better described by the model of People \& Bean.
	
	\begin{figure}[th!]
		\centering
		\includegraphics[scale=0.826]{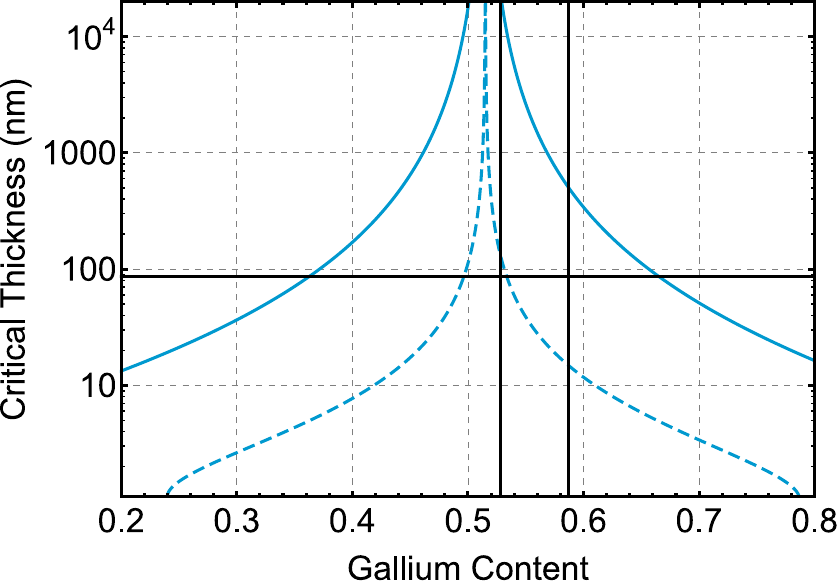}
		\caption{
			Critical thickness, calculated for In$_{1-x}$Ga$_{x}$P on GaAs. Dotted line is the model based on the work of Matthews\cite{Matthews1970JAP-AccommodationOfMisfitAcrossTheInterface} and solid line of People and Bean\cite{People1985APL-CalcHcVersusLatticeMismatchGeSi}. 
			The vertical lines show the Ga contents investigated in this work. 
			The horizontal line indicates the employed thickness 86\,nm.
		}
		\label{SupplementFigure3-CriticalThickness}
	\end{figure}

	\clearpage
	\subsection{High resolution x-ray diffraction measurements}
	
	High resolution x-ray diffraction (HRXRD) (using Cu-K$_{\alpha1}$ radiation) is implemented for the characterization of the structural properties as a non-destructive method with a very high sensitivity to lattice parameter changes.\cite{Holy1996-HRXRDSpaceMapping,Pietsch2004HRXRD} 
	The x-rays scatter from the electronic density of the crystal, reproducing the lattice planes of the crystal. 
	Measuring the angle of the scattered x-rays and using Bragg's law it is possible to calculate the distance between lattice planes, which can be related to the lattice constant of the crystal.
	Thin, mismatched layers distort tetragonally and therefore the lattice parameter parallel to the wafer surface differs from the perpendicular one, $a_\text{L}^\parallel\neq a_\text{L}^\perp$. 
	
	The reciprocal space maps (RSM) were performed using an $\theta$-$\theta$ Empyrean (panalytical) diffractometer equipped with a combination of a Goebel mirror and a single channel cut (Ge220) monochromator on the source side. The scattered intensity was collected using a Pixcel 2D detector (514$\times$514 pixels at 55\,$\mu$m).
	The HRXRD line scans (Fig.\,\ref{SupplementFigure7-HRXRDscan}) were done with a Seifert-GE XRD3003HR diffractometer using a point focus  equipped with an spherical 2D Goebel mirror allowing a beam size in the order of 1\,mm horizontally and vertically. A Bartels monochromator and a triple-axis analyzer in front of a scintillation counter were installed to achieve the highest resolution in reciprocal space.
	
	\subsubsection{HRXRD measurements on high-stress InGaP wafer} \label{suppXRD-HS}
	
	From symmetric RSMs, i.e. the 002 and 004 reflections of Fig. \ref{SupplementFigure4-RSM110} and \ref{SupplementFigure5-RSM1-10}, one can extract the perpendicular lattice constant.
	The following symmetric RSMs show a strong GaAs substrate peak at wave-vectors of about $Q_\text{S[001]}^{002}\approx 2.223$\,\AA$^{-1}$ and $Q_\text{S[001]}^{004}\approx 4.446$\,\AA$^{-1}$, for the 002 and 004 reflection respectively.
	This corresponds to the substrate lattice constant of $a_\text{S}= 5.653$\,\AA.
	The layer peak of InGaP can be seen at $Q_\text{L[001]}^{002}\approx 2.246$\,\AA$^{-1}$ and $Q_\text{L[001]}^{004}\approx 4.493$\,\AA$^{-1}$ and corresponds to the strained, perpendicular lattice constant $a_\text{L}^\perp=5.595$\,\AA.
	
	When additionally measuring asymmetric RSMs, the 224 reflections of Fig. \ref{SupplementFigure4-RSM110} and \ref{SupplementFigure5-RSM1-10} and the 404 reflections of Fig. \ref{SupplementFigure6-RSM404}, it is possible to extract the parallel lattice parameters from the peak position on the $Q_\text{[110]}$ axis. 
	Since both the substrate and layer peak have the same $Q_\text{[110]}$ component, their parallel lattice constants equal: $a_\text{S}=a_\text{L}^\parallel= 5.653$\,\AA, this is true for all the scan directions ([110], $[\overline{1}10]$, [010], $[\overline{1}00]$).
	
	We can clearly see with these findings, $a_\text{S}=a_\text{L}^\parallel \neq a_\text{L}^\perp$, that the high-stress InGaP wafer is 100\,\% pseudomorphic.

	\begin{figure}[th!]
		\centering
		\includegraphics[scale=0.925]{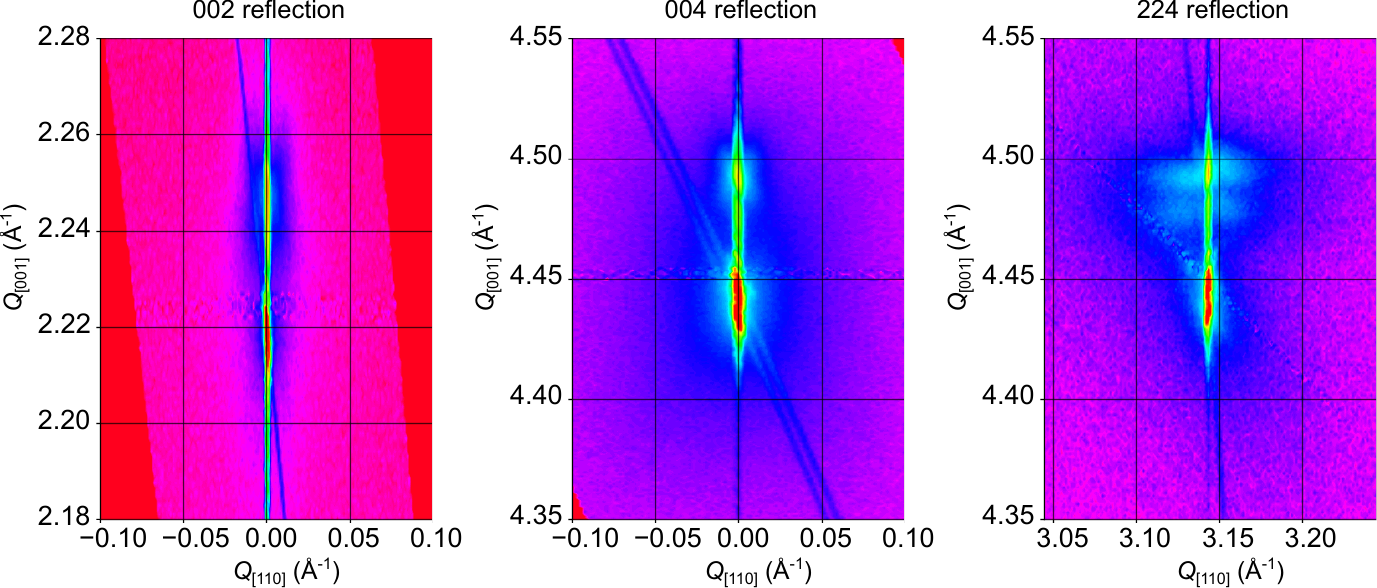}
		\caption{
			HRXRD reciprocal space maps depicting the symmetric 002 and 004 and asymmetric 224 reflections of the high-stress wafer for a x-ray beam oriented along the [110] direction.
		}
		\label{SupplementFigure4-RSM110}
	\end{figure}
	
	\begin{figure}[th!]
		\centering
		\includegraphics[scale=0.925]{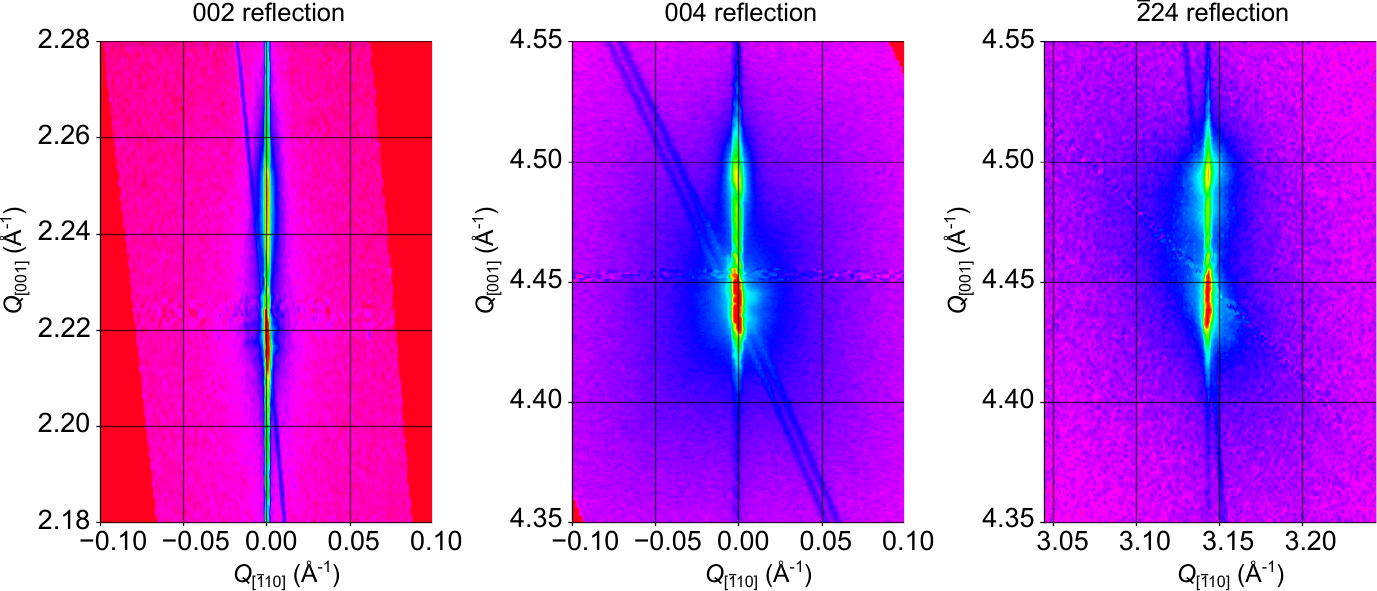}
		\caption{
			HRXRD reciprocal space maps depicting the symmetric 002 and 004 and asymmetric $\overline{2}24$ reflections of the high-stress wafer for a x-ray beam oriented along the $[\overline{1}10]$ direction.
		}
		\label{SupplementFigure5-RSM1-10}
	\end{figure}
	
	\begin{figure}[th!]
		\centering
		\includegraphics[scale=0.925]{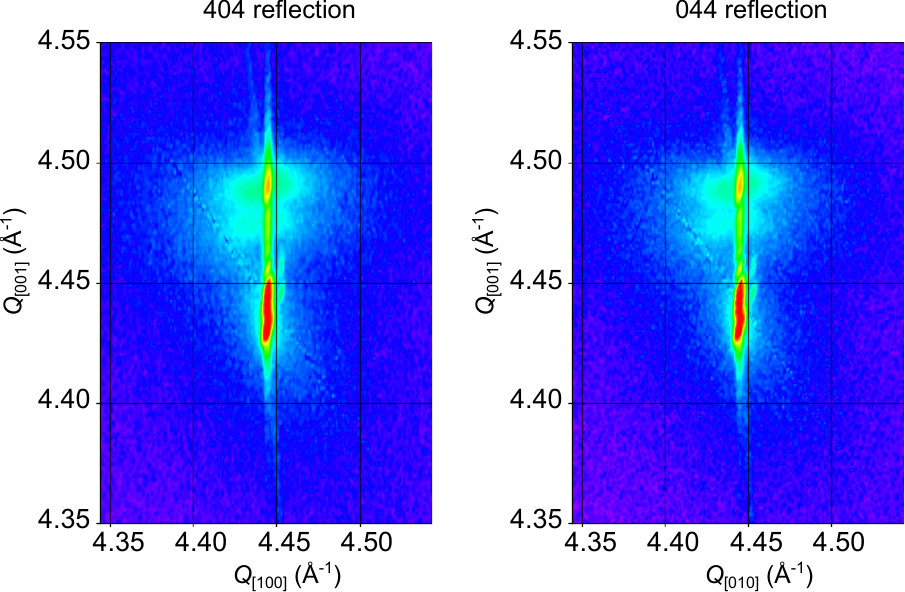}
		\caption{
			HRXRD reciprocal space maps depicting 404 and 044 reflection of the high-stress wafer. On the left the beam is along the $[100]$ and on the right along the [010].
		}
		\label{SupplementFigure6-RSM404}
	\end{figure}

	In addition to the RSMs, to get further insight on the strain, one can look at a reflection curve as a function of scattering angle to extract structural information from the epitaxial structure. 
	Figure \ref{SupplementFigure7-HRXRDscan} shows a HRXRD curve of the 002 reflection in blue.
	The substrate peak can be seen at a scattering angle of about $31.6^\circ$, and the InGaP layer peak at about $32^\circ$. 
	The smaller, regularly distributed peaks arise from the GaAs/AlGaAs super-lattice.
	By simulating and fitting the slow and rapid oscillations of the curve, it is possible to extract the thicknesses and compositions of the different layers in a heterostructure.
	The simulation, red line in Fig. \ref{SupplementFigure7-HRXRDscan}, is done with the heterostructure of Tab. \ref{tabAppendix:ScanInv}.
	The nominal heterostructure composition is shown in Tab. \ref{tabAppendix:Nominal}.
	From the measurement one can see, that the two InGaP layers have different compositions and also point to an In gradient along the growth direction.
	We have not taken into account the compositional gradients in our calculations in the main paper. 
	This will be subject to follow-on work.
	
	\begin{figure}[th!]
		\centering
		\includegraphics{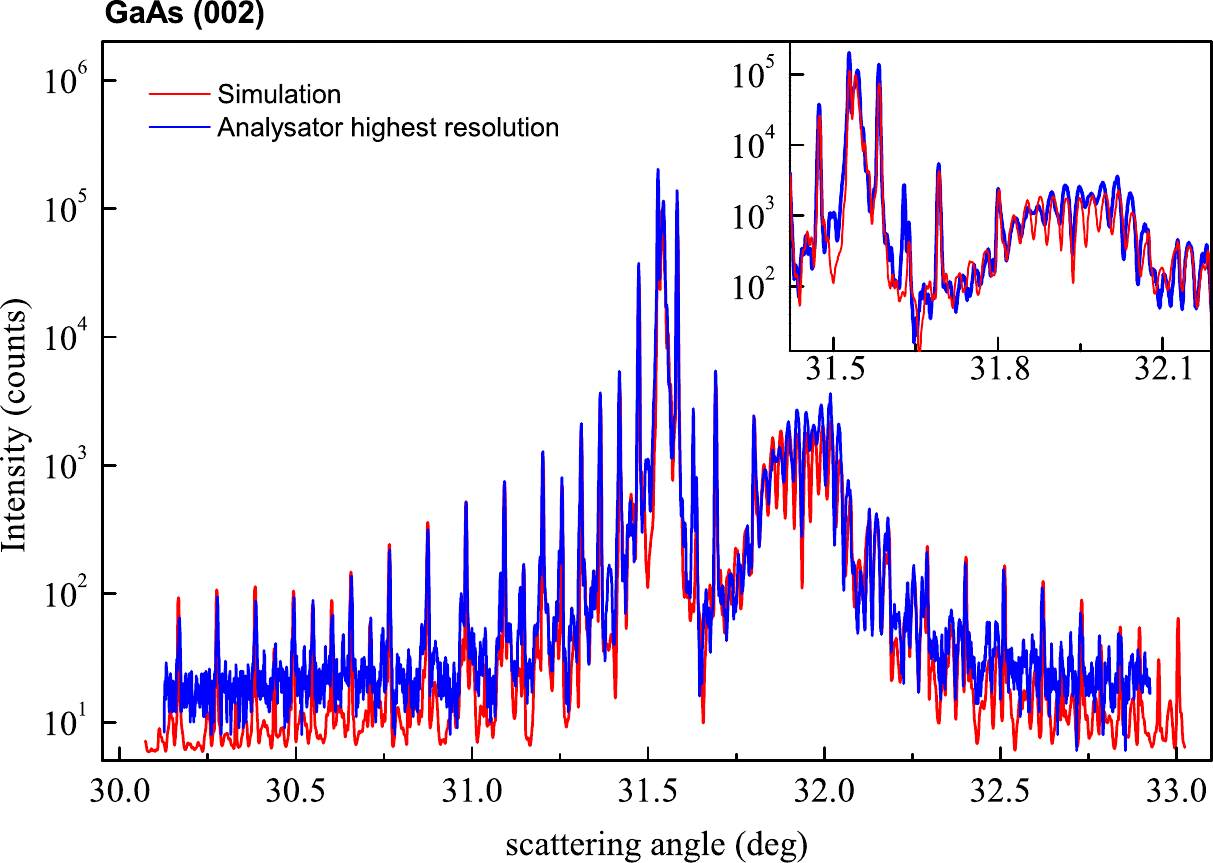}
		\caption{
			HRXRD curve of the 002 reflection for the high-stress InGaP wafer. Fitting the simulation (red) to the signal (blue) leads to the layer-compositions and -thicknesses shown in Table \ref{tabAppendix:ScanInv}.
			Inset: Zoom-in on the InGaP peak at a scattering angle of about 32 degree.
		}
		\label{SupplementFigure7-HRXRDscan}
	\end{figure}

	\begin{table}[th!]
		\caption{
			Heterostructure of the high-stress InGaP wafer. 
			Layer-compositions and -thicknesses are extracted by fitting the HRXRD scan of Fig. \ref{SupplementFigure7-HRXRDscan}.
			One can clearly see deviations from the nominal heterostructure in Tab. \ref{tabAppendix:Nominal}.
			The composition of both InGaP layers differs and they also both have gradient in their In content along the growth direction.
		}
		\begin{tabular}{|l|l|l|l|l|l|}
			\hline 
			Layer & Repeat & Material & Al content & In content & Thickness ($\mu$m) \\ 
			\hline \hline
			0 & 1 & GaAs substrate & --- & --- & --- \\\hline	
			1 & 1 & AlGaAs & 0.99536 & --- & 0.01449 \\\hline
			\multirow{3}{*}{2 } & \multirow{3}{*}{41} & GaAs & --- & --- & 0.07801 \\\cline{3-6}
			& & \multirow{2}{*}{AlGaAs} & 0.96612 (top) & \multirow{2}{*}{---} & \multirow{2}{*}{0.09002} \\
			& &  & 0.98612 (bottom) &  & \\\hline
			3 & 1 & GaAs & --- & --- & 0.07670 \\\hline
			4 & 1 & AlGaAs & 0.96032 & --- & 1.05490\\\hline
			5 & 1 & GaAs & --- & --- & 0.00100 \\\hline
			\multirow{2}{*}{6} & \multirow{2}{*}{1} & \multirow{2}{*}{InGaP} & \multirow{2}{*}{---} & 0.46721 (top) & \multirow{2}{*}{0.07762} \\
			& & & & 0.42233 (bottom) &\\\hline
			7 & 1 & AlGaAs & 0.98550 & --- & 0.26441 \\\hline
			8 & 1 & GaAs & --- & --- & 0.00100 \\\hline 
			\multirow{2}{*}{9} & \multirow{2}{*}{1} & \multirow{2}{*}{InGaP} & \multirow{2}{*}{---} & 0.42227 (top) & \multirow{2}{*}{0.08643} \\
			& & & & 0.38401 (bottom) &\\\hline
			10 & 1 & GaAs & --- & --- & 0.00100 \\\hline
		\end{tabular}
		\label{tabAppendix:ScanInv}
	\end{table}

	\begin{table}[th!]
		\caption{
			Nominal heterostructure composition of the high-stress InGaP wafer.
		}
		\begin{tabular}{|l|l|l|l|l|l|}
			\hline 
			Layer & Repeat & Material & Al content & In content & Thickness ($\mu$m) \\ 
			\hline \hline
			0 & 1 & GaAs substrate & --- & --- & --- \\\hline	
			1 & 1 & AlGaAs & 0.92 & --- & 0.2720 \\\hline
			\multirow{2}{*}{2 } & \multirow{2}{*}{41} & GaAs & --- & --- & 0.0780 \\\cline{3-6}
			& & AlGaAs & 0.92  & --- & 0.0906 \\\hline
			3 & 1 & GaAs & --- & --- & 0.0780 \\\hline
			4 & 1 & AlGaAs & 0.92 & --- & 1.0653\\\hline
			5 & 1 & GaAs & --- & --- & 0.0010 \\\hline
			6 & 1 & InGaP & --- & 0.41  & 0.0859 \\\hline
			7 & 1 & GaAs & --- & --- & 0.0010 \\\hline
			8 & 1 & AlGaAs & 0.92 & --- & 0.2646 \\\hline
			9 & 1 & GaAs & --- & --- & 0.0010 \\\hline 
			10 & 1 & InGaP & --- & 0.41 & 0.0859 \\\hline
			11 & 1 & GaAs & --- & --- & 0.0010 \\\hline
		\end{tabular}
		\label{tabAppendix:Nominal}
	\end{table}

	\subsubsection{HRXRD measurements on low-stress InGaP wafer and comparison to high-stress InGaP}
	
	For the low-stress wafer we performed the same RSM measurements as in subsection \ref{suppXRD-HS}.
	Due to the lower lattice mismatch the InGaP layer peak, at about $Q_\text{L[001]}^{004}\approx 4.452$\,\AA$^{-1}$, is located close to the underlying substrate and AlGaAs super-lattice peak.
	Similar to the high-stress wafer, the InGaP layer peaks show a different diffuse scattering in the asymmetric 224 reflections.
	We clearly see the enhanced diffuse scattering along the [110] crystal direction.

	\begin{figure}[th!]
		\centering
		\includegraphics[scale=0.925]{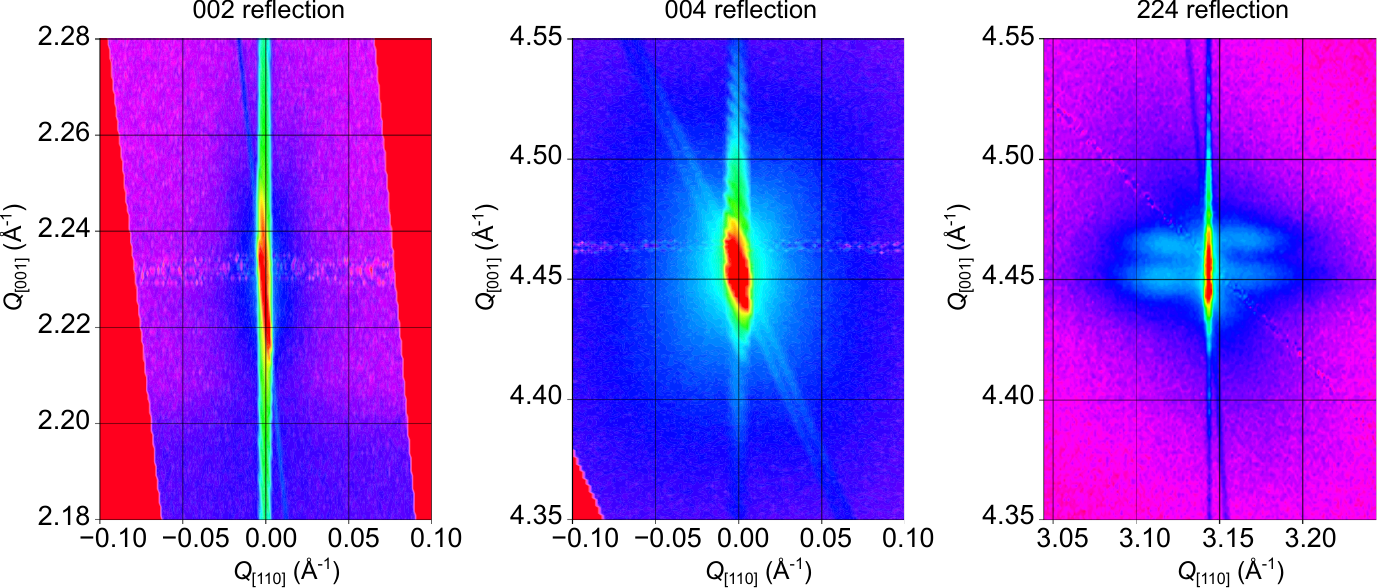}
		\caption{
			HRXRD reciprocal space maps depicting the symmetric 002 and 004 and asymmetric 224 reflections of the low-stress wafer for a x-ray beam oriented along the [110] direction.
		}
		\label{SupplementFigure8-RSM110LS}
	\end{figure}
	
	\begin{figure}[th!]
		\centering
		\includegraphics[scale=0.925]{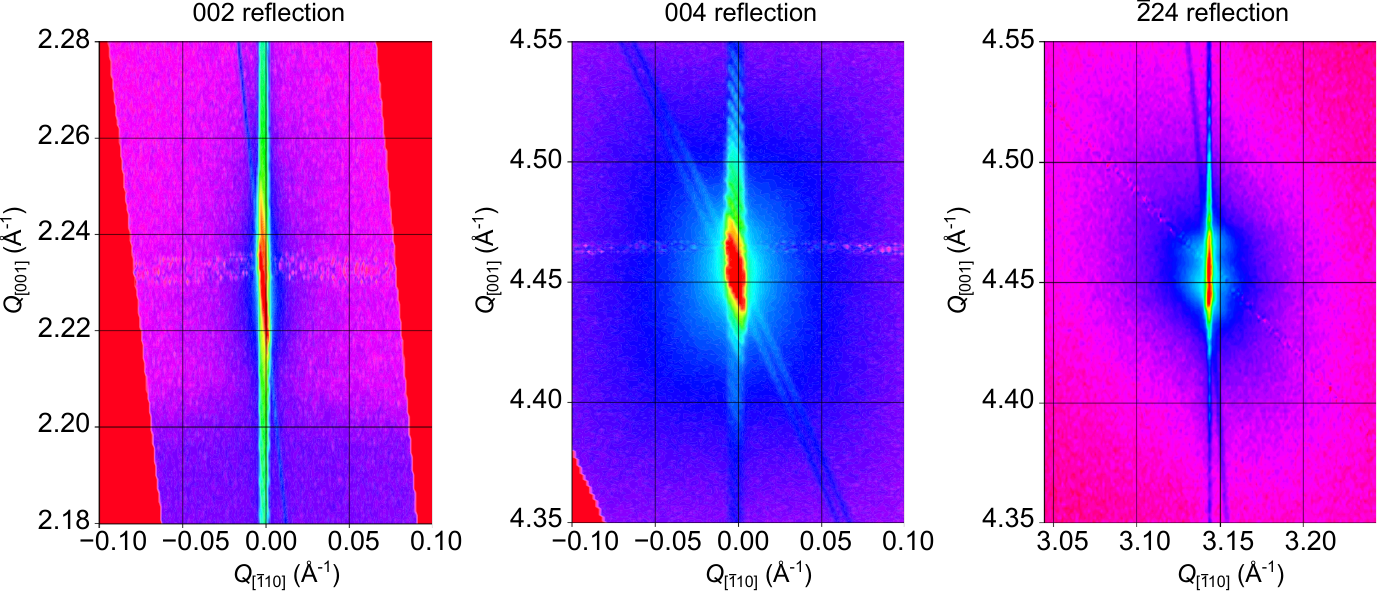}
		\caption{
			HRXRD reciprocal space maps depicting the symmetric 002 and 004 and asymmetric $\overline{2}24$ reflections of the low-stress wafer for a x-ray beam oriented along the $[\overline{1}10]$ direction.
		}
		\label{SupplementFigure9-RSM-110LS}
	\end{figure}
	
	\begin{figure}[th!]
		\centering
		\includegraphics[scale=0.925]{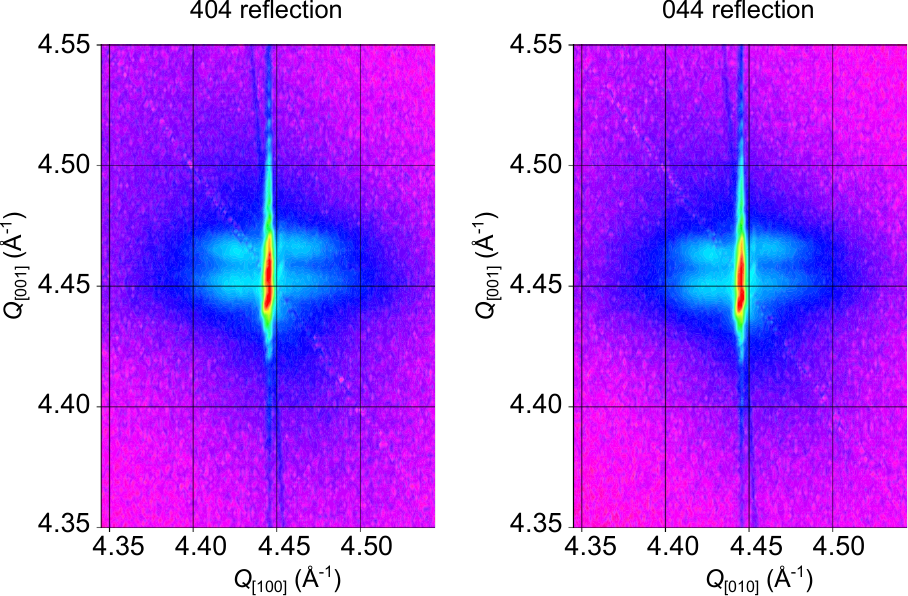}
		\caption{
			HRXRD reciprocal space maps depicting 404 and 044 reflection of the low-stress wafer. On the left the beam is along the $[100]$ and on the right along the [010]
		}
		\label{SupplementFigure10-RSM404LS}
	\end{figure}

	\clearpage
	
\end{widetext}


\end{document}